\documentclass[aps,prl,twocolumn,showpacs,superscriptaddress,groupedaddress]{revtex4}  % for review and submission
\usepackage{graphicx}  % needed for figures
\usepackage{bm}        % for math
\usepackage{amssymb}   % for math
\usepackage{amsmath}   % to split equation
\usepackage{amsfonts}
\usepackage{color}

\begin{document}

%\setpagewiselinenumbers
%\modulolinenumbers[1]
%\linenumbers
\title{Longitudinal Double-Spin Asymmetries for $\pi^{0}$s in the Forward Direction for 510 GeV Polarized $pp$ Collisions}
\affiliation{AGH University of Science and Technology, FPACS, Cracow 30-059, Poland}
\affiliation{Argonne National Laboratory, Argonne, Illinois 60439}
\affiliation{Brookhaven National Laboratory, Upton, New York 11973}
\affiliation{University of California, Berkeley, California 94720}
\affiliation{University of California, Davis, California 95616}
\affiliation{University of California, Los Angeles, California 90095}
\affiliation{Central China Normal University, Wuhan, Hubei 430079}
\affiliation{University of Illinois at Chicago, Chicago, Illinois 60607}
\affiliation{Creighton University, Omaha, Nebraska 68178}
\affiliation{Czech Technical University in Prague, FNSPE, Prague, 115 19, Czech Republic}
\affiliation{Nuclear Physics Institute AS CR, Prague 250 68, Czech Republic}
\affiliation{Technische Universitat Darmstadt, Germany}
\affiliation{Frankfurt Institute for Advanced Studies FIAS, Frankfurt 60438, Germany}
\affiliation{Fudan University, Shanghai, 200433 China}
\affiliation{Institute of Physics, Bhubaneswar 751005, India}
\affiliation{Indiana University, Bloomington, Indiana 47408}
\affiliation{Alikhanov Institute for Theoretical and Experimental Physics, Moscow 117218, Russia}
\affiliation{University of Jammu, Jammu 180001, India}
\affiliation{Joint Institute for Nuclear Research, Dubna, 141 980, Russia}
\affiliation{Kent State University, Kent, Ohio 44242}
\affiliation{University of Kentucky, Lexington, Kentucky 40506-0055}
\affiliation{Lamar University, Physics Department, Beaumont, Texas 77710}
\affiliation{Institute of Modern Physics, Chinese Academy of Sciences, Lanzhou, Gansu 730000}
\affiliation{Lawrence Berkeley National Laboratory, Berkeley, California 94720}
\affiliation{Lehigh University, Bethlehem, Pennsylvania 18015}
\affiliation{Max-Planck-Institut fur Physik, Munich 80805, Germany}
\affiliation{Michigan State University, East Lansing, Michigan 48824}
\affiliation{National Research Nuclear University MEPhI, Moscow 115409, Russia}
\affiliation{National Institute of Science Education and Research, HBNI, Jatni 752050, India}
\affiliation{National Cheng Kung University, Tainan 70101 }
\affiliation{Ohio State University, Columbus, Ohio 43210}
\affiliation{Institute of Nuclear Physics PAN, Cracow 31-342, Poland}
\affiliation{Panjab University, Chandigarh 160014, India}
\affiliation{Pennsylvania State University, University Park, Pennsylvania 16802}
\affiliation{Institute of High Energy Physics, Protvino 142281, Russia}
\affiliation{Purdue University, West Lafayette, Indiana 47907}
\affiliation{Pusan National University, Pusan 46241, Korea}
\affiliation{Rice University, Houston, Texas 77251}
\affiliation{Rutgers University, Piscataway, New Jersey 08854}
\affiliation{Universidade de Sao Paulo, Sao Paulo, Brazil, 05314-970}
\affiliation{University of Science and Technology of China, Hefei, Anhui 230026}
\affiliation{Shandong University, Jinan, Shandong 250100}
\affiliation{Shanghai Institute of Applied Physics, Chinese Academy of Sciences, Shanghai 201800}
\affiliation{State University of New York, Stony Brook, New York 11794}
\affiliation{Temple University, Philadelphia, Pennsylvania 19122}
\affiliation{Texas A\&M University, College Station, Texas 77843}
\affiliation{University of Texas, Austin, Texas 78712}
\affiliation{University of Houston, Houston, Texas 77204}
\affiliation{Tsinghua University, Beijing 100084}
\affiliation{University of Tsukuba, Tsukuba, Ibaraki 305-8571, Japan}
\affiliation{Southern Connecticut State University, New Haven, Connecticut 06515}
\affiliation{University of California, Riverside, California 92521}
\affiliation{University of Heidelberg, Heidelberg, 69120, Germany }
\affiliation{United States Naval Academy, Annapolis, Maryland 21402}
\affiliation{Valparaiso University, Valparaiso, Indiana 46383}
\affiliation{Variable Energy Cyclotron Centre, Kolkata 700064, India}
\affiliation{Warsaw University of Technology, Warsaw 00-661, Poland}
\affiliation{Wayne State University, Detroit, Michigan 48201}
\affiliation{Yale University, New Haven, Connecticut 06520}

\author{J.~Adam}\affiliation{Creighton University, Omaha, Nebraska 68178}
\author{L.~Adamczyk}\affiliation{AGH University of Science and Technology, FPACS, Cracow 30-059, Poland}
\author{J.~R.~Adams}\affiliation{Ohio State University, Columbus, Ohio 43210}
\author{J.~K.~Adkins}\affiliation{University of Kentucky, Lexington, Kentucky 40506-0055}
\author{G.~Agakishiev}\affiliation{Joint Institute for Nuclear Research, Dubna, 141 980, Russia}
\author{M.~M.~Aggarwal}\affiliation{Panjab University, Chandigarh 160014, India}
\author{Z.~Ahammed}\affiliation{Variable Energy Cyclotron Centre, Kolkata 700064, India}
\author{N.~N.~Ajitanand}\affiliation{State University of New York, Stony Brook, New York 11794}
\author{I.~Alekseev}\affiliation{Alikhanov Institute for Theoretical and Experimental Physics, Moscow 117218, Russia}\affiliation{National Research Nuclear University MEPhI, Moscow 115409, Russia}
\author{D.~M.~Anderson}\affiliation{Texas A\&M University, College Station, Texas 77843}
% ...

\collaboration{STAR Collaboration}\noaffiliation
\date{\today}

\begin{abstract}
The STAR Collaboration reports measurements of the longitudinal double-spin asymmetry, $A_{LL}$,
for neutral pions produced at forward directions in polarized proton-proton collisions,
at a center-of-mass energy of $510$ GeV.
Results are given for transverse momenta in the
range $2<p_{T}<10$ GeV/$c$ within two regions of pseudorapidity that span $2.65<\eta<3.9$.
These results are sensitive to the polarized gluon parton distribution function, 
$\Delta g(x)$, down to the region of Bjorken $x \sim 10^{-3}$.
The asymmetries observed are less than $\pm 5 \cdot 10^{-3}$ in magnitude, and
will help constrain the contribution to the spin of the proton from polarized gluons 
at low $x$,
when combined with other measurements as part of a global analysis.
\end{abstract}

\maketitle
%\linenumbers
%\begin{linenumbers}
While Quantum Chromodynamics (QCD) is a highly successful theory of hadronic
interactions, much of its descriptive content must be determined experimentally. 
One such example is the Jaffe-Manohar proton spin sum 
rule $1/2 = \Delta\Sigma/2 + \Delta G + L$,
in which the spin of the proton is expressed as the sum of contributions
from the spins of the 
quarks and antiquarks ($\Delta\Sigma$) and gluons ($\Delta G$), as well as the  
partons' orbital angular momentum ($L$)~\cite{JM}.
The first two terms are integrals
over momentum fraction, $x$, of the
polarized quark and gluon parton distribution functions (PDFs).
Deep Inelastic Scattering (DIS) experiments of charged leptons on polarized targets have fixed the intrinsic quark and antiquark contributions over a wide range of $x$, and give an integral for
$\Delta\Sigma \sim 0.24$~\cite{Sigma1, Sigma2, Sigma3, Sigma4, Sigma5} at a momentum-transfer squared, 
$Q^2 =$10 GeV$^2$/c$^2$. 
This surprisingly small value 
%
%Blumlein & Bottcher  0.193 +/- 0.075
% NNPDF  Q2=1 0.25 +/- 0.09  Q2=10 0.23 +/- 0.16
% DSSV Q2=1 0.255  Q2=10 0.242
%
leaves the origin of the proton
spin largely an unanswered question.  While the \textit{unpolarized} gluon parton
distribution function as a function of $x$ and $Q^2$ can be extracted from scaling violations in
$e-p$ collider data, current facilities for \textit{polarized} DIS studies do not have
sufficient kinematic reach to provide data of comparable quality.

As the world's only polarized proton-proton collider, the Relativistic Heavy Ion Collider (RHIC)
at Brookhaven National Laboratory can uniquely perform spin 
experiments which are complementary to polarized DIS~\cite{RHIC0}.
In proton-proton collisions, signals such as jets and pions are
copiously produced through $gg$, $qg$ and $qq$ hard-scattering processes. 
The corresponding production cross sections for jets
and many hadron species are well-described by global analyses that incorporate experimental data into a
next-to-leading-order (NLO) perturbative QCD theoretical framework~
\cite{NLOXS, Adamczyk:2013yvv, NLOXS1, eun, phenix_cr}
for center-of-mass energies of 200~GeV and above,
from central to forward rapidities, and over a large range of transverse momenta.
These same NLO calculations predict that in hadron production the processes with initial states
$gg$ and $qg$ predominate over $qq$ in the kinematic regions accessible at RHIC, 
giving sensitivity to gluons and leading 
to efficient methods to extract 
the contribution of the polarized gluon PDF from spin asymmetry measurements.

The longitudinal double-spin asymmetry is defined as
\begin{equation}
\label{eqn1}
A_{LL} = \frac{\sigma^{++} - \sigma^{+-} }{ \sigma^{++} + \sigma^{+-} }~,
\end{equation}
where $\sigma^{++}~(\sigma^{+-})$ is the differential pion or jet production cross section for
proton beams with the same (opposite) helicities.
The STAR Collaboration has recently published data for 200 GeV $pp$ collisions on $A_{LL}$ for jets and dijets 
at central pseudorapidity, which are sensitive to $\Delta g(x)$, the polarized gluon distribution function, in the
region $x>0.05$ \cite{NLOXS,Adamczyk:2014ozi, Ting}.
Sensitivity to the gluon polarized PDF in the region $x \sim 0.01-0.05$ has been
explored using $\pi^{0}$s at intermediate pseudorapidities by 
STAR (200 GeV $pp$ collisions~\cite{Adamczyk:2013yvv}), and at mid-rapidity by
PHENIX (510~GeV $pp$ collisions~\cite{phenix_cr}). 
In the present study, we extend this kinematic range to lower $x$ by studying $\pi^{0}$s at forward rapidities.
 
\begin{figure}
\centerline{\includegraphics[scale=0.35]{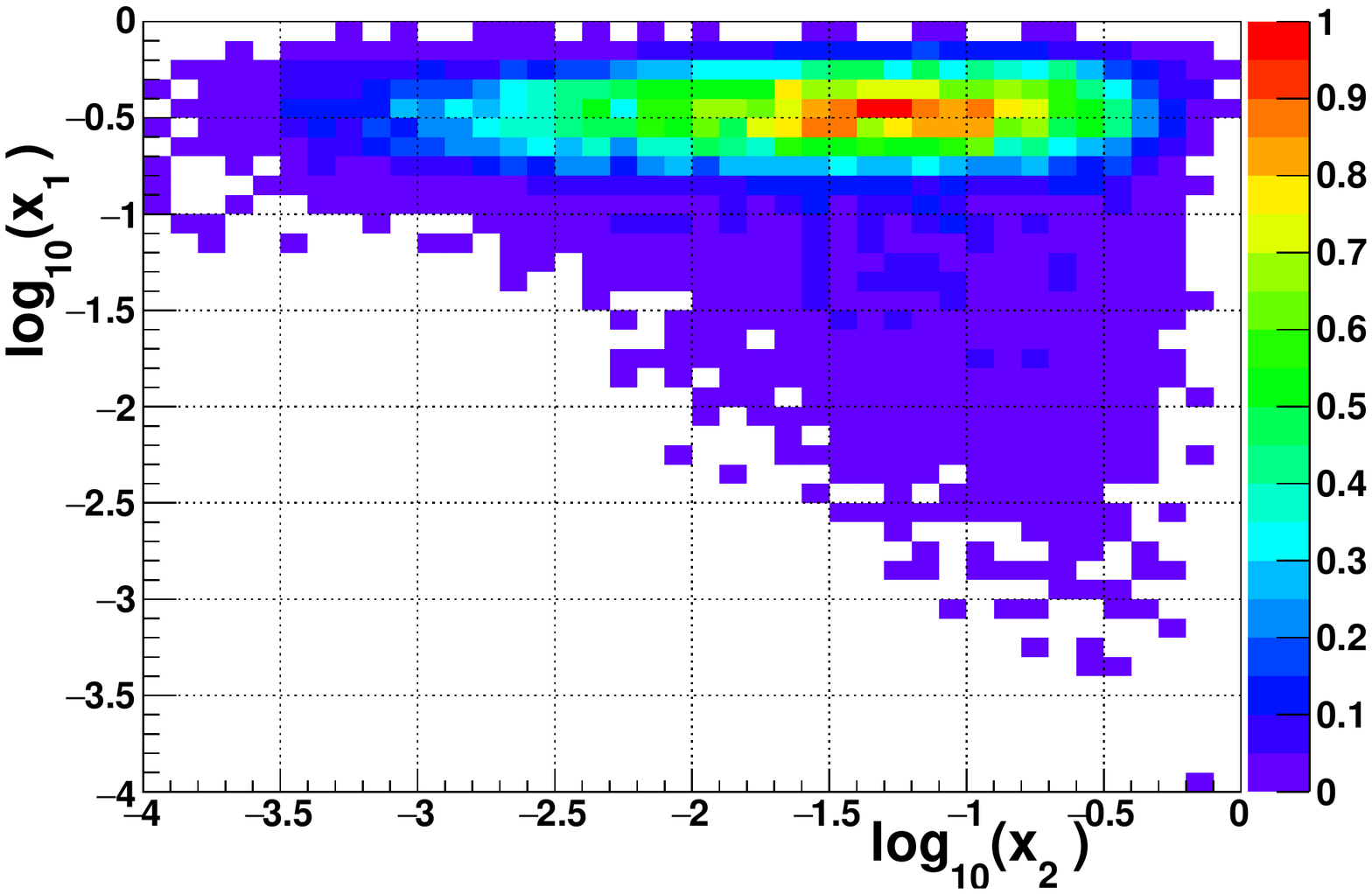}}
\centerline{\includegraphics[scale=0.35]{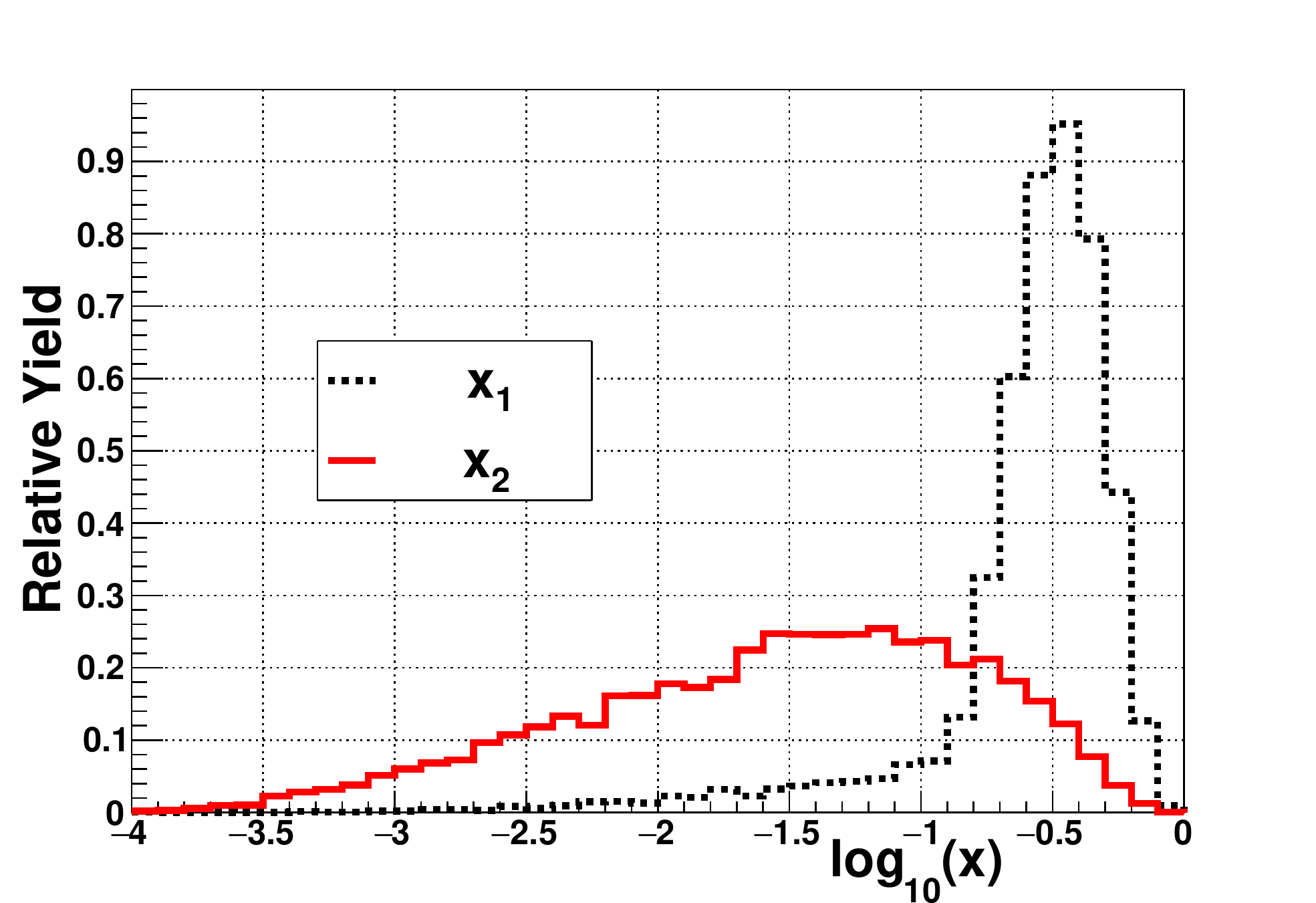}}
\caption{\label{x_fms}Top Panel: Monte Carlo simulations of $x_{1}$ 
\textit{vs} $x_{2}$ for $p p \rightarrow \pi^0 X$ collisions at $\sqrt{s}=510$ GeV.    
The outgoing pion has kinematic cuts in pseudorapidity, transverse momentum and energy of: $2.65<\eta<3.90$, 
$3<p_T<10$ GeV/$c$ and $30<E<70$ GeV, where positive $\eta$ is defined with respect to the direction of 
proton 1 (containing partonic $x_1$), heading into the detector.  
These simulations use PYTHIA Version 6.4.28~\cite{PYTHIA6}, as described in the text.
The scales in both plots are arbitrarily normalized.  Lower Panel: One-dimensional 
projections of $x_{1}$ and $x_{2}$ of the two dimensional histogram.  
}
\end{figure}

This article reports measurements of
$A_{LL}$ for neutral pions in the forward direction,  where 
a pion of longitudinal momentum, $p_L$, carries momentum fraction $2 p_{L}/\sqrt{s}>0.1$. 
While calculations of inclusive particle production cross sections generally involve 
contributions from the underlying PDFs 
over a range of $x$ values, the quasi-two-body nature of the
hard processes and knowledge of the quark polarized PDFs can be used to determine
the range of gluon $x$-sensitivity in a given kinematic range through leading-order (LO) simulations.
In a picture from LO QCD, particles with appreciable transverse momenta are produced from two partons,
one from each of the colliding protons. Forward particles are produced when a 
high-$x$ parton (most likely a quark) in the proton beam moving towards 
the forward detector collides with a 
low-$x$ parton (most likely a gluon) in the proton coming from the detector direction.  
This intuition is confirmed in Fig.~\ref{x_fms} where we present a simulation 
of the range of $x$ sampled
by the two partons using PYTHIA 6.4.28~\cite{PYTHIA6} with the CTEQ6L1~\cite{CTEQ6L1} 
unpolarized PDF set
and the Perugia2012 Parameter Tune~\cite{PERUGIA2012} with the energy-dependence exponent PARP(90) = 0.213.
This tune was selected and adjusted to give the best description of
(unpolarized) charged hadron and jet transverse momenta spectra and 
multiplicities at central rapidities for RHIC data.  
Designating the momentum fraction of partons in the proton which is heading $towards$ the 
forward detector as $x_{1}$ and those
of the proton heading $away$ from the detector as $x_{2}$, then neutral pions
with transverse momentum range $3<p_{T}<10$ GeV/$c$,
energy range $30<E_{\pi^{0}}<70$ GeV and pseudorapidity range $2.65<\eta<3.90$ 
originate from partons with $x_2$ in the range 0.001-0.1.
Because the polarized quark PDFs are already well-determined over the range 
$x_1 > 0.01$~\cite{Sigma1, Sigma2, Sigma3, Sigma4, Sigma5}, 
our asymmetry measurements will be able to help constrain $\Delta g(x)$ down to $x \sim 0.001$. 

The data presented were taken using the Forward Meson Spectrometer (FMS) subsystem of the STAR
{\color{black}experiment~\cite{FMS_bland}}
at RHIC during the years of operation 2012 and 2013. 
The collision energy of $\sqrt{s}=510$ GeV was 
slightly larger than the nominal 500 GeV of previous years in an attempt to improve 
polarized beam operations by using a different operating point and
spin tune for the collider. 
The colliding beams at RHIC are arbitrarily labeled by momentum direction 
as Blue and Yellow: the Blue (Yellow) beam heads toward (away from) the detector, and
hence contains the $x_1$ ($x_2$) parton.

\begin{figure}
\vspace{-0.5cm}
\centerline{\includegraphics[scale=0.40]{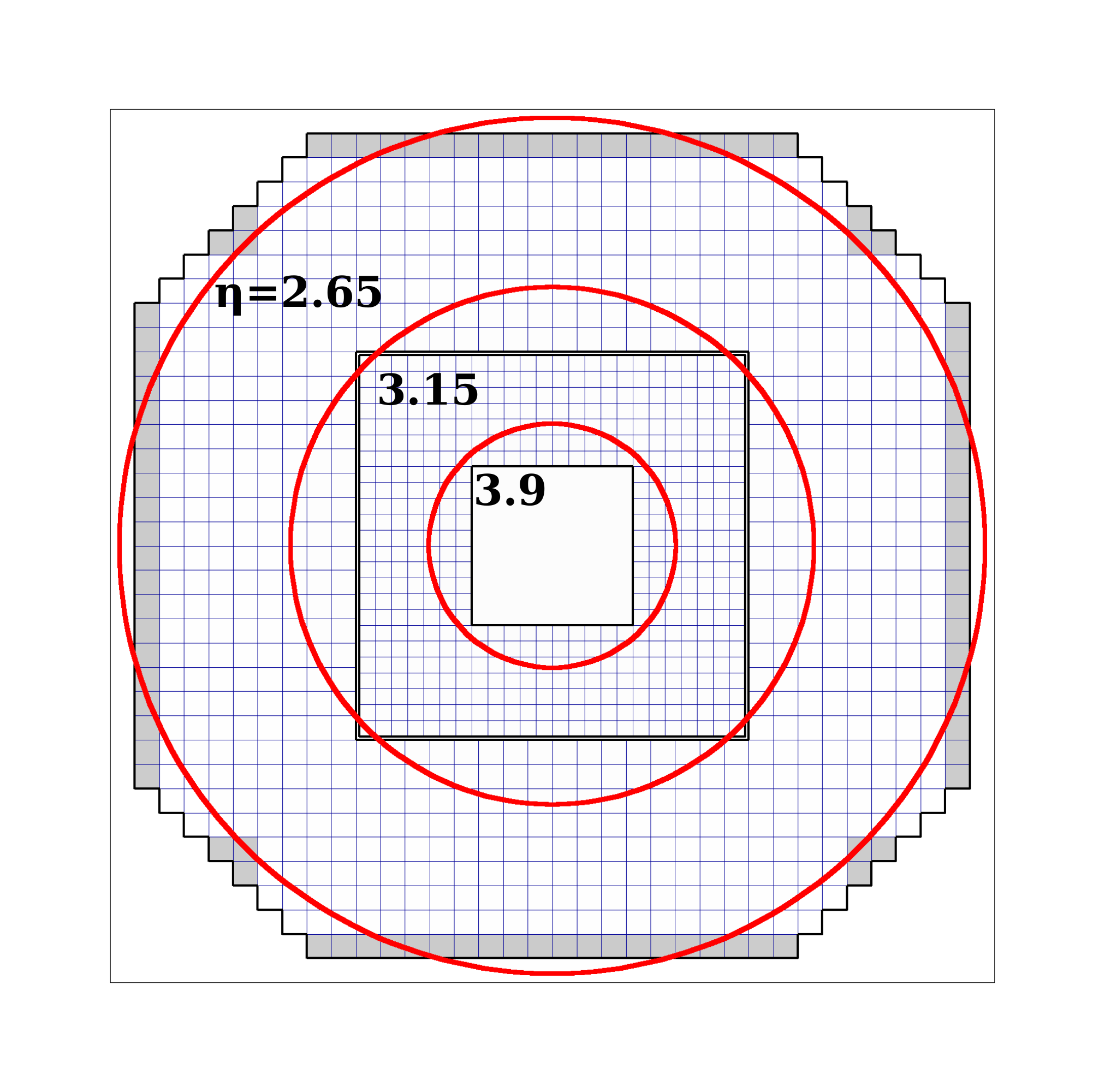}}
\caption{\label{fms}Schematic diagram of the Forward Meson Spectrometer.
The detector is an octagonal arrangement of 788 large and 476 small lead glass 
cells that surround the beam pipe approximately 7~m from the interaction point.  
The shaded cells on the periphery do not participate in the 
definition of the event trigger. Circles are
labeled with values of pseudorapidity cuts used in the analysis, 
which divide the detector into inner and outer regions.
}
\end{figure}

The FMS is a highly-segmented, octagonal wall of lead glass, surrounding the beam pipe with approximately 1~m 
in radius. It is located
7~m from the nominal interaction point of the STAR experiment, in the forward direction of the Blue beam.
A detector schematic is given in Fig.~\ref{fms}. The inner portion consists of a 100~cm $\times$ 100~cm 
square array with a 40~cm $\times$ 40~cm square hole around the beam pipe.  The inner 476 small
cells have dimensions about 3.8~cm $\times$ 3.8~cm $\times$ 45~cm, corresponding to a 
depth of 18 radiation lengths.  
The outer region surrounding the small cells is a set of 788 large cells, 5.8~cm $\times$ 5.8~cm $\times$ 60~cm 
(19 radiation lengths).  The entire array subtends the pseudorapidity range of 
approximately $2.5<\eta<4.0$. The cells are optically isolated from each other using 25 $\mu$m aluminized 
Mylar, and read out by individual photomultiplier tubes which are optically coupled to the lead glass.
The detector is described in further detail in Refs.~\cite{FMS_bland} and \cite{FMS}.

The device is triggered by computing fast sums of the digitized phototube signals in 
regions of different sizes and applying a threshold.  
The first type of trigger, 
the Board Sum (BS), is computed as the sum for overlapping areas corresponding to the 
transverse shower size expected for neutral pions with energies in the region $10-100$ GeV, 
\textit{i.e.}, roughly a patch of 4x8 cells at a distance of 7~m from the interaction region. 
A second type of trigger, the Jet Patch (JP), 
is then formed by grouping these BS regions together into 6 overlapping regions 
each comprising the size of a quarter of the detector.

Both of these triggers consider the transverse energy sum of all the cells in the region.  
For the inner cell BS triggers, we applied $E_T$ thresholds of 1.6 and 2.7 GeV.  
For the outer cell BS triggers, we applied $E_T$ thresholds of 2.9 and 4.3 GeV 
during 2012, and 2.4 and 3.4 GeV during 2013.  For the JP triggers, we applied $E_T$ 
thresholds of 2.8 and 4.3 GeV during 2012, and 1.9 and 3.5 GeV during 2013.  
Generally, the data of each type consist predominantly of events satisfying the higher 
threshold, while events with the lower threshold were pre-scaled due to their larger 
rates and the finite bandwidth of the STAR data acquisition system.

The analysis of an event begins by searching for clusters of contiguous cells with a combined energy
deposition greater than 1~GeV. Since it is expected that the showers 
from the two decay photons for high energy pions will merge when the two 
photon separation becomes comparable to the cell sizes, each cluster of contiguous energy
deposition must be classified as containing one photon or two photons.

Each cluster is characterized using a principal components analysis
method~\cite{wang, eun_thesis, yuxi}.  The log-weighted centroid of the cluster
is determined, and based on this centroid, the covariance matrix elements are
computed.  The larger of the two eigenvalues of this matrix is the first
principal component, which represents the variance of
the cluster along the direction of maximum width, and is a useful parameter for
classifying 1-photon and 2-photon clusters. Plotting distributions of
this quantity in bins of cluster energy reveals two peaks, with the large values 
associated with 2-photon clusters.

After each cluster is categorized, the number, energy and positions of photons 
within the cluster are identified
on the basis of a $\chi^2$ test using a functional form of the transverse shower shape
for one or two photons. This functional form~\cite{lednev} was derived from electron 
test beam data~\cite{FMS_testrun} and isolated photons in
RHIC data~\cite{eun}.
Single photon clusters contain an average of 8 towers over
threshold while a cluster of two (or more) photons contains an average of 12.
This algorithm, which distinguishes between 1-photon and 2-photon clusters, was 
used to extend the useful range of the calorimeter to find
neutral pions with an energy up to 70(100) GeV in the outer(inner) parts of the detector. 
Linear weighting was found to give less discriminating power.
In the kinematic region of data presented in this paper ($p_T < 10$~GeV/$c$), the background
of single cluster contamination in the selected two photon signal is estimated from simulations~\cite{yuxi}
to be less than a few percent of the background under the $\pi^0$ mass peak.

Once the photon candidates have been identified, they are grouped
into cones.  Beginning with the direction of the highest energy photon candidate,
we iteratively search for lower energy photons within a cone of
35 mrad, re-weighting the direction of the cone, until we have 
geometrically divided the event into
a set of cones with photon candidates within those cones.
We then form the invariant mass of the two highest-energy 
photons within the highest-energy cone.
This $\pi^{0}$ candidate is the only $\pi^{0}$ candidate that 
receives further consideration in this event.
Given our cuts on the $p_T$ of the reconstructed $\pi^0$, described in detail below,
this $\pi^0$ will have very likely caused the event trigger.
We have made a cut
on the energy sharing between the two photons of the decay, $z = |E_1 - E_2|/(E_1+E_2) <0.8$.
After computation of the invariant mass, a final cut is made on the transverse
momentum of the pair. This cut varies with time because of different 
PMT calibrations and radiation damage.  The minimum threshold is $p_T>3.0$~GeV/$c$ 
for the outer region of the detector and $p_T>2.0$~GeV/$c$ for the inner region.
Given the upper limits of energy and lower limits of pseudorapidity, the largest
kinematically allowed $p_T$ values are 9.8~GeV/$c$ for the outer and 8.6~GeV/$c$ for
the inner regions of the detector.
While pion yields and backgrounds depend somewhat on the choice of cone size,
the asymmetries are not as sensitive.  The analysis was repeated for a cone size of 100 mrad and the
final results are the same, within statistical errors.

\begin{figure}
\vspace{-0.5cm}
\includegraphics[width=0.50\textwidth]{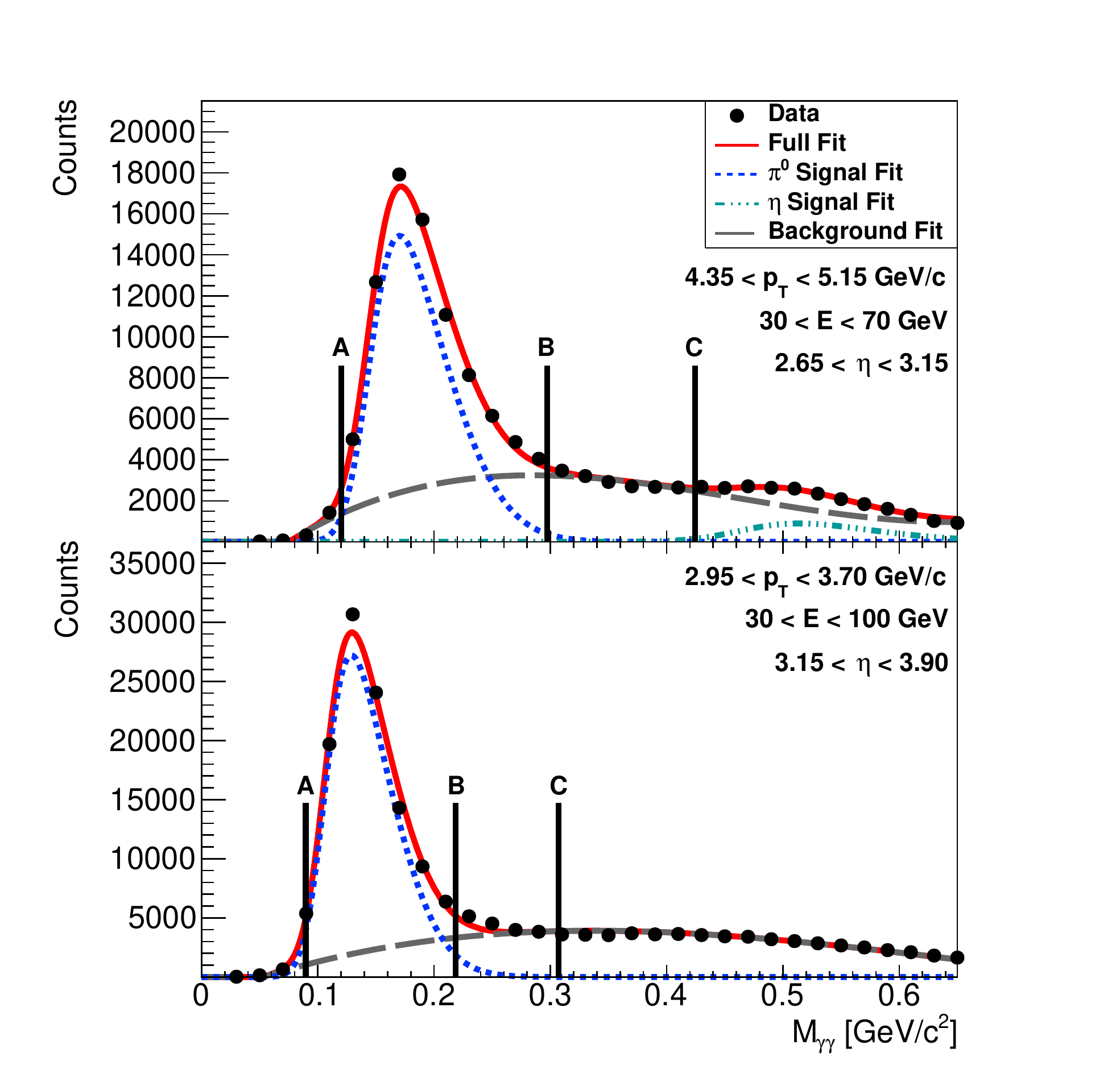}
\caption{\label{p12p13} 
Typical invariant mass spectra for di-photon events
which pass the cuts described in the text. The top plot is for the pseudorapidity range
$2.65<\eta<3.15$ and the bottom plot is for $3.15<\eta<3.90$.
{\color{black} All energy corrections have been applied.
The interval between lines A and B indicate the limits for defining a neutral pion, with a side-band
region defined by the interval BC at larger invariant mass for asymmetry calculations in these bins.   
}
}
\end{figure}

Figure~\ref{p12p13} shows invariant mass distributions of the selected photon pairs, 
with all other $\pi^0$ kinematic cuts applied.  The (large) width of the pion mass
peak is mainly determined by the position resolution of the clusters and
the width of the interaction vertex distribution ($\sim 45$~cm), 
both of which smear the di-photon opening angle.
Simulations of the detector demonstrate that at such high energies the
cluster-finding algorithm generates a decidedly asymmetric shape to
the residual pion signal, after subtraction of background sources such
as falsely-split clusters and combinatorics from photons
from different parent pions~\cite{yuxi}. 
For both data and simulations, the signal shape is found to be
well-fit using a skewed Gaussian for the peak, plus a background
modeled by Chebychev polynomials of degree 3.
For the three lowest $p_T$ bins of the 
outer region, an additional $\eta$-meson signal fit was also 
included. For each $p_T$ bin analyzed for $A_{LL}$, the $\pi^{0}$ 
signal purity was obtained by utilizing the ratio of the 
background fit result, integrated over a signal window determined 
from the skewed Gaussian, to the total number of $\pi^0$ 
candidates in that window. 
Typical background fractions were 10-15$\%$ for the inner region and
20-25$\%$ for the outer region.

A background $A_{LL}$ value was determined from the sideband 
invariant mass region of the photon pairs, between 
the $\pi^0$ and $\eta$-meson signal regions, 
in order to correct for a possible background asymmetry contribution.
This method assumes the background has the same $A_{LL}$ value within the $\pi^0$ peak region as within
the sideband region at higher invariant mass.
Given the background $A_{LL}$ and the $\pi^0$ signal 
purity for each $p_T$ bin, the $\pi^0$ signal $A_{LL}$ 
was extracted from the $\pi^0$+background $A_{LL}$.

There was a significant degradation 
in resolution between the years 2012 and 2013 
due to darkening of the lead glass by radiation damage.
This darkening causes a decrease in the light output of the glass,
which has been accounted for by re-calibrating the detector every
few days using the centroids of the pion and eta mass peaks.  
The radiation damage was worse for the inner cells, especially close to the beam pipe,
so we only included outer cell data in the analysis of the 2013 data.

Because the longitudinal profile of radiation damage leads to a 
corresponding decrease in transparency of the lead glass,
this causes a change in the effective position of the shower maximum, which is
used in the estimate of the position and direction of the primary energetic photon.
The reconstructed pion mass shifts to larger values with increasing energy, as the
determination of the energy and opening angles of the two decay photons becomes less
accurate.  
To compensate for this effect, we define the mass range for pions in the analysis 
using an energy-dependent mass window, plus a side-band to determine the background 
as indicated by the vertical lines in Fig.~\ref{p12p13}.  
This window is adjusted to contain the reconstructed pion peak over the entire kinematic range
of the spin asymmetry measurements.

Data were aquired in short runs of 10-60 minutes.
After expressing the cross sections in Eq.~\ref{eqn1} in terms of quantities measured on a run-by-run basis,
we have, by the maximum likelihood method,
\begin{equation}
\label{eqn2}
A_{LL} = \frac{ \Sigma(P_B P_Y)(N^{++} - r N^{+-})} { \Sigma(P_B P_Y)^{2}(N^{++} + r N^{+-})} ~,
\end{equation}
where $P_{B,Y}$ are the polarization values for each beam, $N^{++}~$($N^{+-}$) are the
inclusive pion yields for beams of the same (opposite) helicities, and the relative luminosity, 
$r \equiv \mathcal{L}^{++}/\mathcal{L}^{+-}$, is the ratio of the luminosities for 
bunches with each beam helicity combination. 
The summations are taken over runs, where for each run
the yields for different helicity combinations and relative luminosities are 
computed. These measurements are then combined 
with polarization measurements, which are taken at regular intervals throughout each fill.
The RHIC rings are loaded with beams 
having 111 bunches circulating in opposite directions, with polarization fill patterns
constructed to reduce possible systematic correlations between polarization and
bunch number in RHIC or the STAR detector~\cite{RHICpol}. 

Spin-dependent pion yields in Eq.~\ref{eqn2} are measured by sorting bunch combinations 
during a run, resulting in a suppression of systematic errors
due to secular variations in the detector efficiency or beam conditions.
The polarization of
the beams is measured at the beginning of, end of, and every 3 hours during a beam store
using dedicated polarimeters based on proton-Carbon scattering in the Coulomb-nuclear interference
region~\cite{CNI}, and calibrated against a polarized atomic hydrogen gas-jet target~\cite{gasjet}.

The relative luminosities are measured on a run-by-run basis. 
For this purpose, STAR is equipped with several sets of detectors in
different ranges of pseudorapidity, which are sensitive to
different physics processes, beam background conditions and
absolute counting rates. For these measurements, we used the Vertex Position Detectors (VPD)~\cite{VPD}
which are a pair of Pb convertor/scintillation counters,
each with 19 segments, located $\pm5.7$~m ($4.24<|\eta|<5.21$) from the nominal interaction point.
As an independent measurement of the relative luminosity, we employed the STAR 
Zero Degree Calorimeters (ZDC)~\cite{ZDC}
which are a pair of tungsten-plate/PMMA-fiber-ribbon calorimeters designed to be sensitive to neutrons,
situated between the RHIC rings at a distance of $\pm18$~m from the interaction point. 
Counts from these detectors were directed to a 30-bit,
redundant scaler system which incremented every 106.5~ns beam bunch-crossing for each data run.
A 7-bit identifier was allocated for bunch-crossing number in order to determine the spin combination
for each scaler count. 

Although beam/background conditions, luminosity and detector performance differed significantly
for the two years of data-taking, the estimation of the systematic error on the relative luminosity
gave similar results. In 2012 (2013) the relative luminosity ratios for bunches with same/opposite-sign 
helicities were in the range $0.94-1.06$ ($0.92-1.08$).  Careful inter-comparison of pairs of detectors
and scaler systems revealed that the most reliable consistency was to be found between
the VPD and ZDC and gave a systematic uncertainty in $A_{LL}$ of approximately $3\cdot10^{-4}$ 
due to the relative luminosities for both years. Because two detectors could be used to
measure the relative luminosity, the systematic uncertainty is defined by how well the
relative luminosity measurements agree with each other. 
Three methods were used to assess this agreement: 
(1) a comparison of the relative luminosity measurement between
the VPD and ZDC, (2) a bias from a possible double-spin asymmetry in the VPD or ZDC themselves, and
(3) an evaluation of the transverse single-spin asymmetry seen in the VPD. While method (3) involves only the
VPD, it helps validate methods (1) and (2) by providing an independent assessment of the impact of relative
luminosity uncertainty on a spin asymmetry. Ultimately, all three measurements of the
relative luminosity systematic uncertainty are in agreement.

\begin{figure}
\vspace{-0.5cm}
\includegraphics[width=0.50\textwidth]{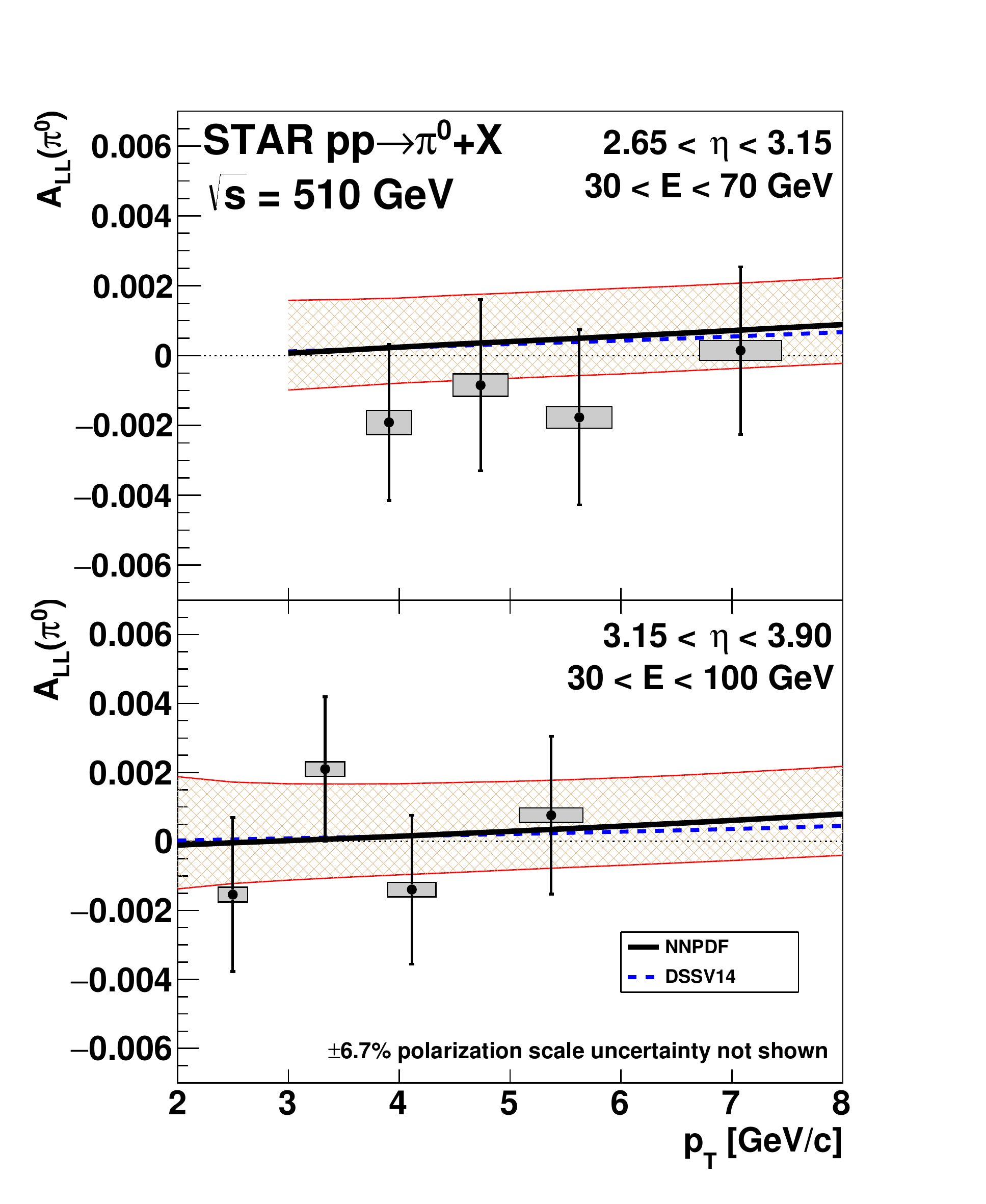}

\caption{\label{all_vs_pt1} Longitudinal Double-Spin Asymmetry, $A_{LL}$ \textit{vs} $\pi^{0}$ transverse momentum
in polarized $pp$ collisions at $\sqrt{s}=510$ GeV in the pseudorapidity (energy) 
ranges $2.65<\eta<3.15$ ($30<E_{\pi}<70$ GeV) (top) and  $3.15<\eta<3.90$ ($30<E_{\pi}<100$ GeV) (bottom).  
Data collected in 2012 and 2013 have been combined.
Vertical error bars on the data represent the statistical uncertainties from pion
yields and polarization measurements only.  
The vertical extent of the shaded boxes gives the combined systematic uncertainties from the 
relative luminosity and polarization measurements.
Measurements of the
beam polarization give a multiplicative uncertainty on these data due to
the factor $1/P_{Y} P_{B}$ equal to $\pm6.7\%$~\cite{rhicspin}, which is not shown.
The horizontal extent of the shaded boxes represent the $p_T$ systematic uncertainty, described in the text.
On the same graphs we plot theoretical calculations of $A_{LL}$ for neutral pions~\cite{JAGER}, 
using the NNPDFpol1.1~\cite{NNPDF1.1} (black solid line and error band for the 100 replicas in the set) and DSSV14~\cite{DSSV14}
(blue dashed line) 
sets of polarized PDFs.  In both cases, we use the DSS fragmentation functions~\cite{DSS}.  
}
\end{figure}

The spin asymmetries are calculated using a maximum likelihood method that 
weights each event according to the relative luminosity in each run and the polarization
in each fill and sums these quantities over the course of the entire data-taking period.  
The 2012+2013 data have a combined luminosity of about 63~pb$^{-1}$ and an average polarization
of $54.6\pm1.9\%$ in the Blue Beam and $56.4\pm2.0\%$ for the Yellow Beam.
The measured $A_{LL}$ points are plotted in Fig.~\ref{all_vs_pt1} for two different ranges of
pseudorapidity of the pion.  The asymmetry values are plotted at the mean transverse
momenta of each bin. The vertical error bars represent the statistical
errors, calculable from the pion yields and polarization measurements on the data.
The vertical extent of the gray boxes gives the uncertainties on $A_{LL}$ values
arising from systematic uncertainties on the relative luminosities and possible remnant
transverse components of beam polarization in the RHIC machine.
The horizontal extent of the gray boxes represent the $p_T$ systematic uncertainties, 
which were approximately 5.2$\%$.  
The energy calibration uncertainty makes the dominant
contribution, since the precision of the energy 
calibration is estimated to be $\pm5\%$ at pion energies in the range of 
$20-40$~GeV. 

Accounting for correlations of the errors on the polarization in each beam gives a relative
error on the product $1/P_{Y}P_{B}$ of $\pm$6.7$\%$ for the combined 2012+2013 run periods~\cite{rhicspin}.
This error should be considered as an overall vertical scale uncertainty on the data, but is omitted
for clarity in the plots.

While the dominant systematic errors on $A_{LL}$ were those associated with the relative luminosities
and beam polarization measurements, many other sources of systematic error 
were considered and estimated.
One contribution to an apparent longitudinal double-spin asymmetry could arise from the
residual transverse components of the beam polarization (typically about $5\%$ of the longitudinal
component), in conjunction
with the $transverse$ double asymmetry $A_{\Sigma}$ as defined in Ref.~\cite{rathmann}.
Measurements at 500 GeV of $A_{\Sigma}$ as a function of pion $p_T$ give results 
which are consistent with zero. As in previous STAR
longitudinal double-spin asymmetry measurements~\cite{Adamczyk:2012qj}, we did not make a 
correction to $A_{LL}$, but instead assigned a conservative systematic 
uncertainty to the $A_{LL}$ measurements to account for a possible correction.
We estimated this by combining the measurements of $A_{\Sigma}$ with measurements 
of the transverse polarization
components of the Blue (Yellow) beams.
These contributions to the
systematic errors on $A_{LL}$ are found
to be on the order of $10^{-5}$ and are, thus, negligible compared to the systematic error due to the
relative luminosity and polarization measurements. 

The longitudinal double-spin asymmetry of jets and neutral pions gives sensitivity to $\Delta g$, 
since the associated cross sections are dominated by gluonic subprocesses and the 
PDFs for polarized quarks and antiquarks are known with comparatively much greater 
precision~\cite{Sigma1, Sigma2, Sigma3, Sigma4, Sigma5}. 
Recent measurements of the longitudinal double-spin asymmetry for inclusive
jets at central rapidity in STAR~\cite{Adamczyk:2014ozi} have been incorporated into
global analyses~\cite{NNPDF1.1, DSSV14} and suggest that the
integral of $\Delta g(x,Q^{2}=10~$GeV$^{2}$/c$^{2})$
over the range $0.05<x<1.0$ is positive~\cite{nocera},
with the two analyses giving consistent values of $0.23\pm0.06$~\cite{NNPDF1.1} and
$0.20+0.06/-0.07$~\cite{DSSV14}.
To determine the net gluon spin contribution to the proton,
assumptions must be made about the shape of the polarized gluon 
PDF in the unmeasured regions,
especially at low $x$, where it is poorly constrained by the existing data.
In Fig.~\ref{all_vs_pt1} we have plotted the
predictions for $A_{LL}$ using a NLO calculation~\cite{JAGER}
but substituting the PDF sets
from NNPDFpol1.1~\cite{NNPDF1.1}/NNPDF2.3~\cite{NNPDF1} and DSSV14~\cite{DSSV14}/CTEQ6M~\cite{CTEQ6L1}.
The presented data points are consistent with both of these extrapolations of $A_{LL}$ to
these kinematic ranges.
In each case
we have used the DSS fragmentation functions~\cite{DSS}.  Both of these fits include RHIC
central rapidity data for jets~\cite{Adamczyk:2012qj, Adamczyk:2014ozi, nnpdf28}, while the NNPDFpol1.1 fit includes 
RHIC $W^{\pm}$ data~\cite{nnpdf31, nnpdf32, nnpdf33} as well.  
The error bands for the NNPDF asymmetries in Fig.~\ref{all_vs_pt1} were determined by taking the 100 replicas
of the set and computing the variance of the polarized gluon PDF sampled for each $x$ and $Q^2$ 
used in determining the polarized cross section for a particular pion transverse momentum.  This variance was
then added and subtracted to the central value and the cross section and asymmetry were recomputed.
The sensitivity of these predictions to the renormalization, factorization
and fragmentation scales was checked and found to be negligible, compared to the errors on the data.

The STAR collaboration has also published
data on neutral pion spin asymmetries in the intermediate region ($0.8<\eta<2$)~\cite{Adamczyk:2013yvv},
which are sensitive to the polarized gluon PDF in 
the range $0.01<x<0.05$.  With the present
data, we push the sensitivity for $\Delta g(x)$ to $x \sim 0.001$.
To date, global analyses have only been able to constrain the gluon polarization down to $x \sim $0.01 through extrapolation from the higher $x$ region. These measurements will provide the first direct experimental constraints on $\Delta g(x)$ in this important low-$x$ range.

We thank the RHIC Operations Group and RCF at BNL, the NERSC Center at LBNL, 
and the Open Science Grid consortium for providing resources and support. 
This work was supported in part by the Office of Nuclear Physics within the 
U.S. DOE Office of Science, the U.S. National Science Foundation, the Ministry of Education 
and Science of the Russian Federation, National Natural Science Foundation of China, 
Chinese Academy of Science, the Ministry of Science and Technology of China 
and the Chinese Ministry of Education, the National Research Foundation of Korea, 
GA and MSMT of the Czech Republic, Department of Atomic Energy and Department of 
Science and Technology of the Government of India; the National Science Centre of Poland, 
National Research Foundation, the Ministry of Science, Education and Sports of the 
Republic of Croatia, RosAtom of Russia and German Bundesministerium fur Bildung, 
Wissenschaft, Forschung and Technologie (BMBF) and the Helmholtz Association.
   % input acknowledgement

%\end{linenumbers}
\end{document}